\documentclass[twocolumn,amsmath,amssymb,showpacs,nofootinbib,prfluids,aps,superscriptaddress]{revtex4-2} 

\bibstyle{apsrev}
\usepackage{bm}%
\usepackage{graphicx}
\usepackage{tikz}
\usepackage{ulem}
\usepackage{color}
\usepackage{xcolor}
\usepackage{physics}
\usepackage[colorlinks=true,linkcolor=blue]{hyperref}%
\expandafter\ifx\csname package@font\endcsname\relax\else
 \expandafter\expandafter
 \expandafter\usepackage
 \expandafter\expandafter
 \expandafter{\csname package@font\endcsname}%
\fi
\begin{document}

\definecolor{pyblue}{HTML}{1F77B4}
\definecolor{pyorange}{HTML}{FF7F0C}
\definecolor{pygreen}{HTML}{2CA02C}
\definecolor{pyred}{HTML}{D62728}

\title{To coalesce or not to coalesce: Droplets and surface tension gradients}

\author{S. Zitz}
\email{zitz@ruc.dk}
 \affiliation{IMFUFA, Department of Science and Environment,\\ Roskilde University, Postbox 260, DK-4000 Roskilde, Denmark}%
 \affiliation{Department of Chemical and Biological Engineering, Friedrich-Alexander-Universit\"at Erlangen-N\"urnberg, F\"{u}rther Stra{\ss}e 248, 90429 N\"{u}rnberg, Germany}%
 \author{T. Richter}%
 \affiliation{Helmholtz Institute Erlangen-N\"urnberg for Renewable Energy,\\
  Forschungszentrum J\"ulich,
  F\"urther Strasse 248, 90429 N\"urnberg, Germany}%
  \affiliation{Department of Physics, Friedrich-Alexander-Universität Erlangen-Nürnberg, \\Fürther Straße 248, 90429 Nürnberg, Germany}%
 \author{K. Missios}%
 \affiliation{IMFUFA, Department of Science and Environment,\\ Roskilde University, Postbox 260, DK-4000 Roskilde, Denmark}%
 \author{J. Roenby}%
\email{johan@ruc.dk}
 \affiliation{IMFUFA, Department of Science and Environment,\\ Roskilde University, Postbox 260, DK-4000 Roskilde, Denmark}%
\date{\today}

\begin{abstract}
We numerically study the coalescence dynamics of two sessile droplets with radii $R_0$.
The droplets are placed on top of a rigid substrate with a contact angle of $\theta_{eq.} = \pi/9$. 
Having a highly wettable substrate ($\theta_{eq} \ll \pi/2$) theory predicts that the bridge height ($h_0$) scales according to $h_0(t) \sim t^{2/3}.$
This behavior can be altered with e.g. surface tension gradients ($\partial_x\gamma \neq 0$). 
These gradients appear for example with heat transfer, surfactants or having different but miscible liquids.
Instead of coalescence, these gradients can lead to a stable two droplet state. 
In this work, we focus on two aspects of this problem.
The first one is the concrete choice of the surface tension, therefore making it spatially correlated.
The second one is the reduction of scale towards a regime in which the disjoining pressure becomes important. 
We find that coalescence can be suppressed, given that there is a sharp gradient in surface tension.
If this gradient is smeared, we find an intermediate agreement with the $2/3$ power-law.
In the limit of large smearing width, we observe an asymmetric coalescence.
\end{abstract}

\maketitle
 
\newcommand{\ts}{\textsuperscript}

\section{Introduction}\label{sec:intro}
Liquid droplets are an ubiquitous phenomenon which has raised scientific curiosity for decades.
Among sprays, sliding drops or sessile drops, there is a vast zoo of different hydrodynamic problems.
The coalescence of liquid droplets or the lack of it is one of these problems. 
Both scenarios can be observed in many instances of our everyday life and in various industrial processes.
Coalescence of sessile droplets happens, for example, when vapor condenses.
During this process, small droplets nucleate randomly and quickly assemble in larger drops, either naturally~\cite{PhysRevA.43.1906} or due to artificial factors such as surface structure or patterning~\cite{C1SM06219K}. 
This effect can effectively be used in so-called fog harvesting devices, where ambient water vapor is condensed and collected in structures which promote droplet coarsening~\cite{zhang2015inkjet, shi2018fog}.
Apart from fog harvesting, coalescence or rather the control of it plays an important part in inkjet printing and printable electronics~\cite{jo2009evaluation, singh2010inkjet, Kim_2005, Luechinger_2008}, but also in various microfluidic devices for mixing purposes at low Reynolds numbers ($Re < 1$)~\cite{https://doi.org/10.1002/pen.760352206, doi:10.1063/1.858199}. 

On the other hand, there are many applications that require droplets not to coalesce.
In hot summer days, a fine water spray can help the body to cool down.
Single, small droplets advect heat from the body and transfer it into the surrounding fluid (air)~\cite{kim2007spray}.
Apart from cooling, emulsions are an illustrative example of systems where droplets should not coalesce.
Mayonnaise is an emulsion that is created by vigorously stirring an oil, egg yolk mixture. 
The stirring breaks up the oil phase in small dispersed droplets.
Proteins and additives like mustard stabilize this state and keep the oil droplets from coalescing, which has a significant effect on the rheology of this mixture~\cite{harrison1985factors, DEPREE2001157}.

More generally, droplet coalescence has attracted much attention in the past two decades, see refs.~\cite{eggers_lister_stone_1999, duchemin_eggers_josserand_2003, PhysRevLett.95.164503, PhysRevLett.106.114501, doi:10.1063/1.4828721}. 
The driving forces during coalescence are, the minimization of surface area and the minimization of interfacial curvature. 
These arguments hold true for scenarios beyond simple Newtonian liquids, in fact they are applicable to liquid lens coalescence~\cite{PhysRevLett.124.194502} or coalescence of quasi 2D liquids~\cite{klopp2020self, doi:10.1021/acs.langmuir.0c02139}.
What is common to the above-mentioned examples is the equilibrium state, which is one where the two droplets have coalesced.
This is of course in agreement with a minimized surface area and minimal curvature.

However, this reasoning doesn't always apply. 
Recently, Kern et al. have shown that viscoplastic effects can arrest the coalescence, yielding a stable twin drop states~\cite{PhysRevFluids.7.L081601}.
What has been known for some time, however, is that a surface tension gradient influences drop coalescence, as demonstrated by Riegler and Lazar and later Karpitschka et al.~\cite{PhysRevLett.109.066103, doi:10.1021/la500459v, karpitschka2014sharp, bruning2018delayed}.
They performed experiments with two drops from different but miscible liquids, therefore having a sharp surface tension gradient from one drop to the other.
Bocia and Bestehorn showed non-coalescence of droplets with numerical simulations~\cite{PhysRevE.82.036312, borcia2011coalescence}.
The clean formulation and derivation was later done by Karpitschka et al.~\cite{PhysRevLett.109.066103}, as they identify an effective Marangoni flow that stabilizes the two droplet system.
We revisit that problem with numerical simulations using a lattice Boltzmann method that is based on the thin film equation~\cite{PhysRevE.100.033313}.
First, we move to the regime where the droplets become tiny and the disjoining pressure can no longer be neglected.
Second, we address the impact of the surface tension beyond the reduction to its absolute contrast.
We ask the straightforward question, if the concrete function of $\gamma(x)$ can turn non-coalescing states into coalescing ones.
Within this framework, we discuss the growth law of the liquid bridge and the influence of the disjoining pressure on coalescence.

This paper is organized as follows:
Starting in the next section, Sec.~\ref{sec:theory}, we discuss the underlying theoretical model.
Introducing the thin film equation with an effective Marangoni contribution. 
For the thin film pressure, we include a disjoining pressure functional $\Pi(h)$, which consists of thickness dependent power-law and a wettability component. 
In Sec.~\ref{sec:method} we discuss the numerical method we use to solve the thin film equation.
We show how to construct an additional term to account for the flow due to the Marangoni effect, and
introduce the three functions we use for the spatially resolved surface tension $\gamma(x)$. 
The results are presented and discussed in Sec.~\ref{sec:results}.
Showing, first, the evolution of the bridge for the various choice of $\gamma(x)$.
Knowing how the bridge evolves, we associate this data with three different scenarios.
Revealing that for small scales, the disjoining pressure can not be neglected.
Finally, we supply a brief summary with conclusion and future research in Sec.~\ref{sec:sum_conclu}. 

\section{Theory}\label{sec:theory}
The theoretical approach we use to study this system is the lubrication approximation~\cite{Reynolds, RevModPhys.69.931, PhysRevE.63.011208}.
Applying this approximation to the Navier-Stokes equation yields the thin film equation, which for a singular horizontal dimension reads~\cite{RevModPhys.81.739, RevModPhys.81.1131, THIELE2014399}
\begin{equation}\label{eq:thin_film_simple}
    \partial_t h = \partial_x \left(\frac{h^3}{3\mu}\partial_x p\right),
\end{equation}
where $\frac{h^3}{3\mu}$ is the mobility $M(h)$\footnote{$m(h) = h^3/3\mu$ is true only for a no-slip velocity boundary condition.}, $h(x,t)$ is the thickness of the film at time $t$ and position $x$, $\mu$ is the liquid's viscosity, and $p$ it's pressure.
The pressure accounts for both the surface tension and the correct fluid substrate behavior (wetting)~\cite{PhysRevE.100.033313}.
We therefore write the pressure as
\begin{equation}\label{eq:pressure}
    p = \gamma \partial_x^2 h + \Pi(h),
\end{equation}
where the first term is the (1D) Laplacian of the liquid-vapor interface and $\Pi(h)$ is the disjoining pressure~\cite{RevModPhys.69.931, RevModPhys.81.739, Peschka9275, PhysRevE.63.011208},
\begin{equation}\label{eq:disjoin}
    \Pi(h) = K(\gamma,\theta)\left[\left(\frac{h_{\ast}}{h}\right)^n - \left(\frac{h_{\ast}}{h}\right)^m\right].
\end{equation}
The prefactor $K(\gamma,\theta)\propto \gamma(1-\cos(\theta))$ encodes the wettability and as such the equilibrium contact angle $\theta_{\text{eq.}}$.
This prefactor can be understood as a linking factor to the Hamaker constant ($\mathcal{A}$) with~~\cite{PhysRevE.93.013120, bestehorn20033d, van1988interfacial}
\begin{equation}
    \mathcal{A} = 6\pi h_{\ast}^3 K(\theta).    
\end{equation}
Additionally, $h_{\ast}$ defines the thickness of the precursor layer, therefore the thickness where $\Pi(h_{\ast}) = 0$.
The pair of powers $(n,m)$ need to satisfy $n > m$ and $m > 1$.
For an in deep presentation of the derivation we refer to the work by Schwartz and Eley and Bonn et al.~\cite{SCHWARTZ1998173, RevModPhys.81.739}.

\begin{figure}
    \centering
    \includegraphics[width=0.48\textwidth]{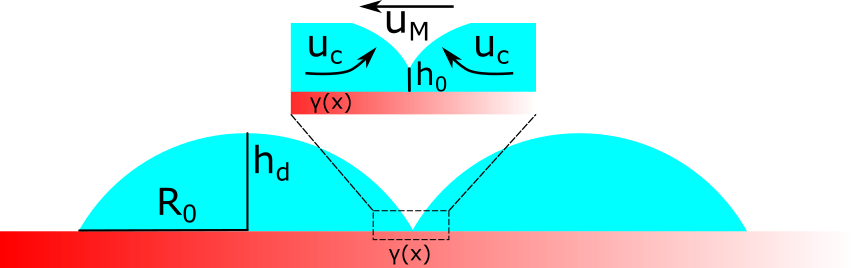}
    \caption{Schematic setup of the numerical experiment. 
    Two droplets with base radii $R_0$ and maximum height $h_d$ are placed next to each other. 
    They are connected by a liquid bridge with height $h_0$ and subject to a surface tension gradient $\gamma(x)$ (red color gradient).
    The flow has two sizeable contributions, one due to capillarity, $u_c$, and one due to Marangoni, $u_M$.
    }
    \label{fig:schematics}
\end{figure}
In the presence of a surface tension gradient, Eq.~(\ref{eq:thin_film_simple}) requires another term.
This term accounts for an effective Marangoni flow~\cite{doi:10.1021/la500459v, karpitschka2014sharp, bestehorn20033d, doi:10.1021/la960488a}
\begin{equation}\label{eq:thin_with_marangoni}
    \partial_t h = \partial_x \left(\frac{h^3}{3\mu}\partial_x p + \frac{h^2}{2\mu}\partial_x\gamma\right).
\end{equation}
These gradients appear due to non-homogenized surfactant concentrations or spatially resolved heating profiles, e.g. with lasers~\cite{doi:10.1021/la960488a, NIKOLOV2002325, bruning2018delayed, wedershoven2014infrared}. 
They appear as well when two different but miscible liquids come into contact with each other~\cite{doi:10.1021/la800630w, karpitschka2014sharp, doi:10.1021/la500459v}. 

\subsection{Flows}\label{subsec:flows_theory}
Eq.~(\ref{eq:thin_with_marangoni}) defines the dynamics of the system we are interested in.
Before dealing with the full complexity, we take a step back and set $\partial_x\gamma = 0$.
In this case, we know that the two touching droplets will coalesce.
Following the argumentation of Riegler, Lazar and Eddie et al.~\cite{doi:10.1021/la800630w, PhysRevLett.111.144502},
\begin{equation}\label{eq:cap_pressure}
    P_{\text{cap.}} \sim\gamma\kappa \sim \frac{\gamma}{h_0},
\end{equation}
where $P_{\text{cap.}}$ is the capillary pressure and $\kappa$ is the curvature of the liquid vapor interface.
Upon rescaling, Eddi et al. found in their analytical analysis that $h_0$ is the only relevant scale, justifying therefore $\kappa \sim 1/h_0$~\cite{PhysRevLett.111.144502}. 
The resulting flow towards the bridge induces an inertial or ``Bernoulli'' pressure
\begin{equation}\label{eq:P_Bernulli}
    P_{\text{iner.}} \sim \rho\left(\frac{h_0}{t}\right)^2.
\end{equation}
Balancing Eqs.~(\ref{eq:cap_pressure}-\ref{eq:P_Bernulli}) and solving for $h_0$ yields a growth law 
\begin{equation}\label{eq:coal_powerlaw}
    h_0(t) \sim t^{2/3},
\end{equation}
for contact angles below $\theta < \pi/2$~\cite{PhysRevLett.111.144502, keller2002breaking}.
Recently, a similar growth law for the coalescence of low viscosity liquid lenses has been experimentally observed and theoretically validated, see Hack et al.~\cite{PhysRevLett.124.194502}.

The addition of a surface tension gradient alters the pressure balance.
For two different but miscible liquids, Riegler and Lazar associated Eq.~(\ref{eq:P_Bernulli}) with the difference in surface tension~\cite{doi:10.1021/la800630w}
\begin{equation}\label{eq:P_bernulli_riegler}
    P_{iner.} \sim \frac{\rho (\delta\gamma)^2}{\mu^2}, 
\end{equation}
where $\delta\gamma = \gamma_1 - \gamma_2$ and $\gamma_i$ are the two surface tensions of two different liquids. 
Karptischka and Riegler later formulated an elegant theoretical explanation for the flow of the non-coalescing twin droplet state~\cite{PhysRevLett.109.066103}.
In their derivation, they assume a quasistatic thickness ($\partial_t h \approx 0$) and a constant motion proportional to the capillary number 
\begin{equation}\label{eq:sys_cap_vel}
    Ca = \frac{\mu v}{\gamma},
\end{equation} 
with $v$ being a characteristic velocity that preserves the bridge's profile. 
The starting point is Eq.~(\ref{eq:thin_with_marangoni}) in the quasistatic limit, 
\begin{equation}\label{eq:pressure_noncoal}
    \partial_x\left(\frac{h^3}{3\mu}\partial_x p + \frac{h^2}{2\mu}\partial_x\gamma\right) = 0.
\end{equation}
Integrating the above equation and isolating the pressure gradient $\partial_x p = \gamma\partial_x^3 h$ yields~\cite{RevModPhys.69.931, PhysRevLett.109.066103}
\begin{equation}\label{eq:karpitschka_h3}
    \partial_x^3 h = \frac{3}{h^2}Ca - \frac{3}{2h\gamma}\partial_x\gamma.
\end{equation}
In the absence of a surface tension gradient we have $\partial_x^3 h \geq 0$.
With a surface tension gradient, however, $\partial_x^3 h$ can change sign~\cite{PhysRevLett.109.066103}.

Our model in addition to the above derivation uses a pressure given by Eq.~(\ref{eq:pressure}), therefore modifying Eq.~(\ref{eq:karpitschka_h3}) slightly, 
\begin{equation}\label{eq:stefan_h3}
    \partial_x^3 h \approx \frac{3}{h^2}Ca - \frac{3}{2h\gamma}\partial_x\gamma - \partial_x\Pi(h).
\end{equation}
where the derivative of the disjoining pressure appears. 
The velocity field inside the droplets (film) can be computed using $\partial_x^3 h$~\cite{RevModPhys.69.931},
\begin{equation}\label{eq:Oron_correct}
    u(z) = \frac{z}{\mu}\left[\partial_x\gamma - \left(\frac{z}{2} - h\right)\gamma\partial_x^3 h\right].
\end{equation}
We therefore assume that the derivative of the disjoining pressure has an effect on the flow, at least in the region close to the bridge.

\section{Method}\label{sec:method}
\begin{figure}
    \centering
    \includegraphics[width=0.48\textwidth]{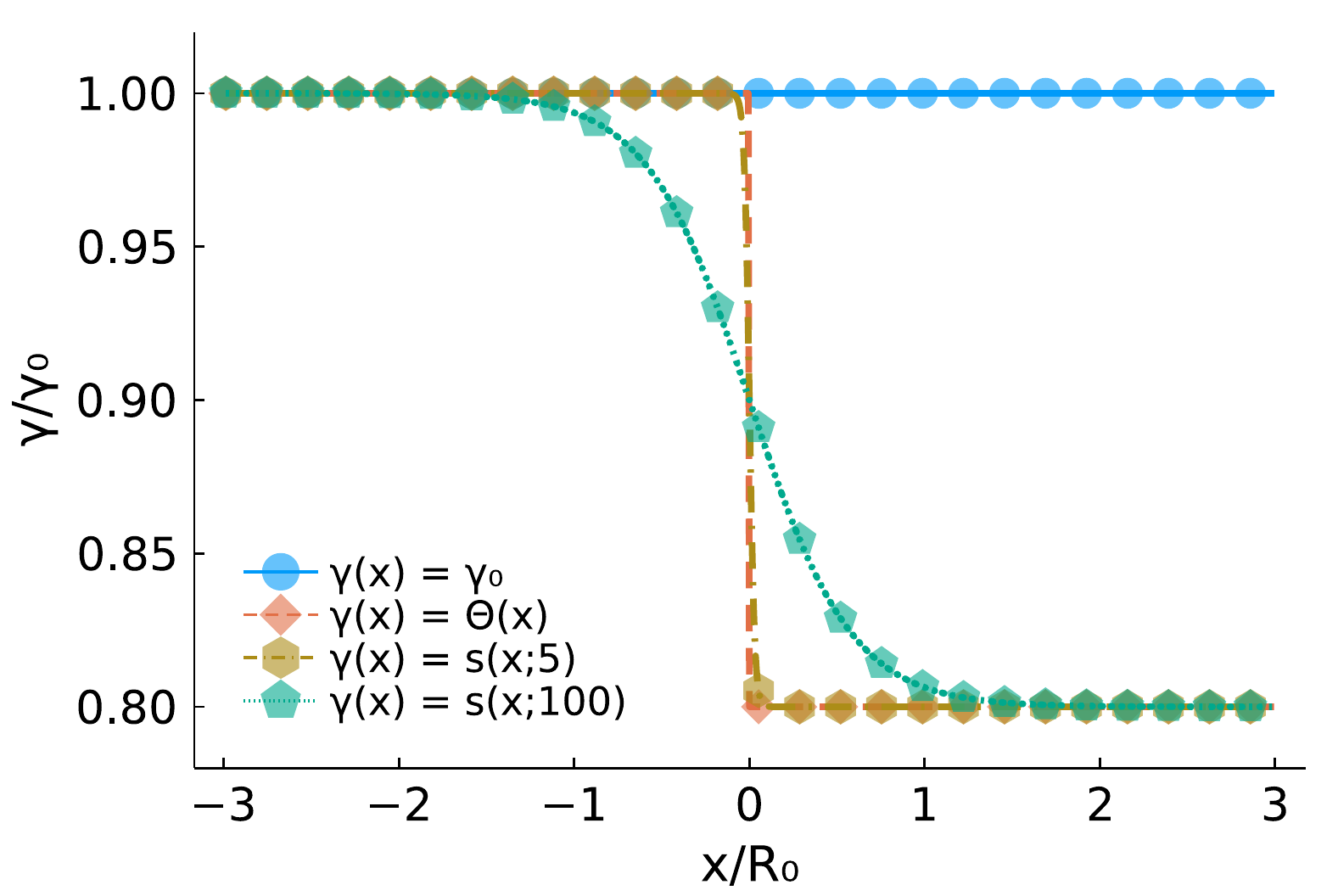}
    \caption{Surface tension fields $\gamma(x)$ normalized with $\gamma_0$.
    Different colors and symbols display different surface tensions with $x=0$ is the coordinate of the bridge at $t=0$.
    The surface tension functions are given by Eqs.~(\ref{eq:gamma_const}-\ref{eq:gamma_tanh}).
    }
    \label{fig:gammas}
\end{figure}
We use a computational method, namely the lattice Boltzmann method (LBM), to iteratively solves Eq.~(\ref{eq:thin_with_marangoni}).
The details of the method can be found in the App.~\ref{app:two} and in the references~\cite{PhysRevE.100.033313, PhysRevE.104.034801}.

In addition to the basic lattice Boltzmann algorithm, we require and another force term $F_{\gamma}$ to account for the Marangoni contribution.
Similarly to earlier work on thermal fluctuations~\cite{PhysRevE.104.034801} we construct a force term and match it to the last term in Eq.~\ref{eq:thin_with_marangoni}, 
\begin{equation}\label{eq:force_gamma_grad}
    F_{\gamma} = \frac{3}{2}\partial_x\gamma(x),
\end{equation}
where we have assumed that the surface tension only varies along the horizontal dimension.
With the addition of this force term, our LBM is a numerical algorithm for the system
\begin{equation}\label{eq:lubr2eq1surf}
\begin{cases}
\begin{array}{ll}
\partial_t h + \partial_x (h u)  = 0 & \\ 
\partial_t (h u) = -\frac{1}{\rho_0}h\partial_x p -\nu\alpha_{\delta}(h)u + \frac{3}{2}\partial_x\gamma.
\end{array}
\end{cases}
\end{equation}

Performing the limits discussed in ref.~\cite{PhysRevE.100.033313, PhysRevE.104.034801}, this system becomes an effective solver for Eq~(\ref{eq:thin_with_marangoni}).
Due to the inclusion of $\Pi(h)$, this method allows us to test the non-coalescence criteria for smaller scales, well below $1mm$ droplet diameter~\cite{doi:10.1021/la500459v, karpitschka2014sharp}.

For the spatially resolved surface tension, $\gamma(x)$ we use three different functions, as shown in Fig.~\ref{fig:gammas}.
The first one is a constant surface tension, 
\begin{equation}\label{eq:gamma_const}
    \gamma(x) = \gamma_0,
\end{equation}
with $\gamma_0$ as constant and therefore $\partial_x\gamma(x) = 0$, see blue curve Fig~\ref{fig:gammas}.
Second, to mimic a mixture of different liquids or a locally heated and cooled substrate, we use a Heaviside function $\Theta(x)$
\begin{equation}\label{eq:gamma_step}
    \gamma(x) = \Theta(x) = \begin{cases}
    \gamma_0\quad~~\qquad \text{for $x < L/2$}\\
    \gamma_0 - \Delta\gamma \quad \text{for $x \ge L/2$}\\
    \end{cases},
\end{equation}
with $\Delta\gamma$ is a percentage of $\gamma_0$, given by the orange dashed curve in Fig.~\ref{fig:gammas}.
The third function interpolates smoothly between the two values $\gamma_0$ and $\gamma_0 -\Delta\gamma$ using a tangent hyperbolicus
\begin{align}\label{eq:gamma_tanh}
    \gamma(x) &= \gamma_0\abs{1 - \left(\frac{1}{2} - s(x;l,w)\right)} + \nonumber\\
    &(\gamma_0 - \Delta\gamma)\left\{1 - \abs{1 - \left(\frac{1}{2} - s(x;l,w)\right)}\right\} 
\end{align}
where
\begin{equation}\label{eq:smoothing}
    s(x;l,w) = \frac{1}{2}\tanh\left(\frac{x - l}{w}\right),
\end{equation}
with the coordinate $l$ where $\gamma(l) = \gamma_0 -\Delta\gamma/2$ and a smoothing width $w$. 
In the following, we keep $l=L/2$ fixed and use the notation $s(x;w)$.
The influence of $w$ is shown in Fig.~\ref{fig:gammas} using Eq.~(\ref{eq:gamma_tanh}) twice with $w=5$ (yellow dashdotted curve) and $w=100$ (cyan dotted curve). 

The coupling of Eqs.~(\ref{eq:gamma_const}-\ref{eq:gamma_tanh}) to the LBM is twofold.
The surface tension is evaluated during the pressure calculation Eqs.~(\ref{eq:pressure}-\ref{eq:disjoin}) and for the Marangoni contribution, Eq.~(\ref{eq:force_gamma_grad}).

\section{Results}\label{sec:results}
\begin{figure*}
    \centering
    \includegraphics[width=1.0\textwidth]{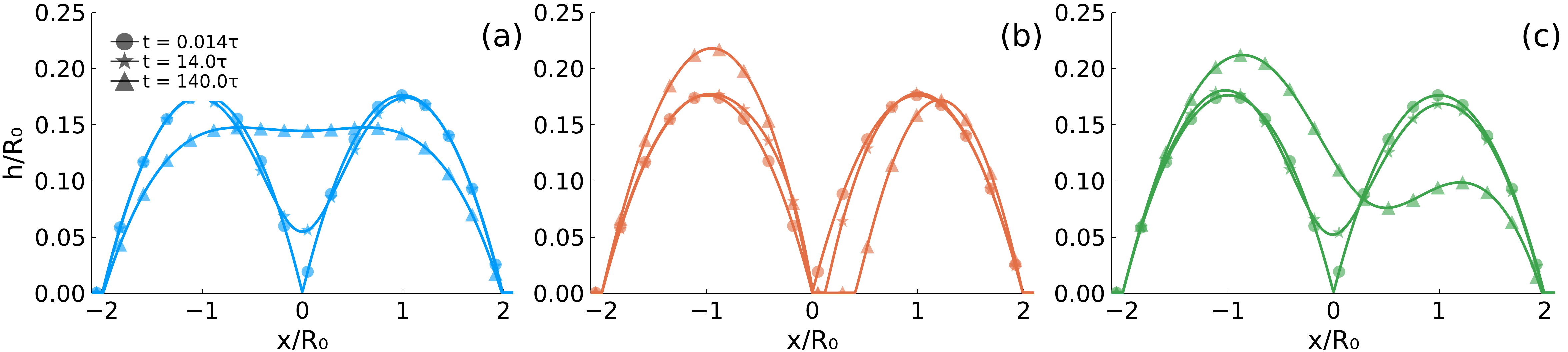}
    \caption{Snapshots of the thickness $h(x)$ taken at three different time steps.
    The three time step have different marks, namely bullets, stars and triangles for $t=0.14\tau$, $t=14\tau$ and $t=140\tau$ respectively.
    In panel (a) we show the evolution for a constant surface tension, see Fig.~\ref{fig:gammas} blue bullets.
    In (b) and (c) we use Eq.~(\ref{eq:gamma_tanh}) with $w =5$ and $w=100$ respectively. 
    }
    \label{fig:final_state}
\end{figure*}
All numerical experiments start with the same initial condition, two barely overlapping circular segments (equivalent to spherical caps in three dimensions) as illustrated in Fig.~\ref{fig:schematics}.
We use the radius of the initial droplets $R_0$ to normalize length scales~\cite{PhysRevLett.111.144502, PhysRevLett.95.164503}.
As a characteristic time scale we use the inertio-capillary time, 
\begin{equation}\label{eq:inertio-cap-time}
    \tau = \sqrt{\frac{\rho R_0^3}{\gamma}}.
\end{equation}

In Fig.~\ref{fig:final_state} we show a time series of our numerical experiments for three different surface tensions.
In blue, we have the case of a constant surface tension, thus a pure coalescence.
The symbols, bullets, stars and triangles display the thickness at $t=0.014\tau$, $t=14\tau$ and $t=140\tau$ respectively. 
The influence of $\gamma(x)$ is best seen by the green and orange curves, as both of them use Eq.~(\ref{eq:gamma_tanh}) but have different $w$ values.

\subsection{Bridge growth}\label{subsec:growth}
\begin{figure}
    \centering
    \includegraphics[width=0.48\textwidth]{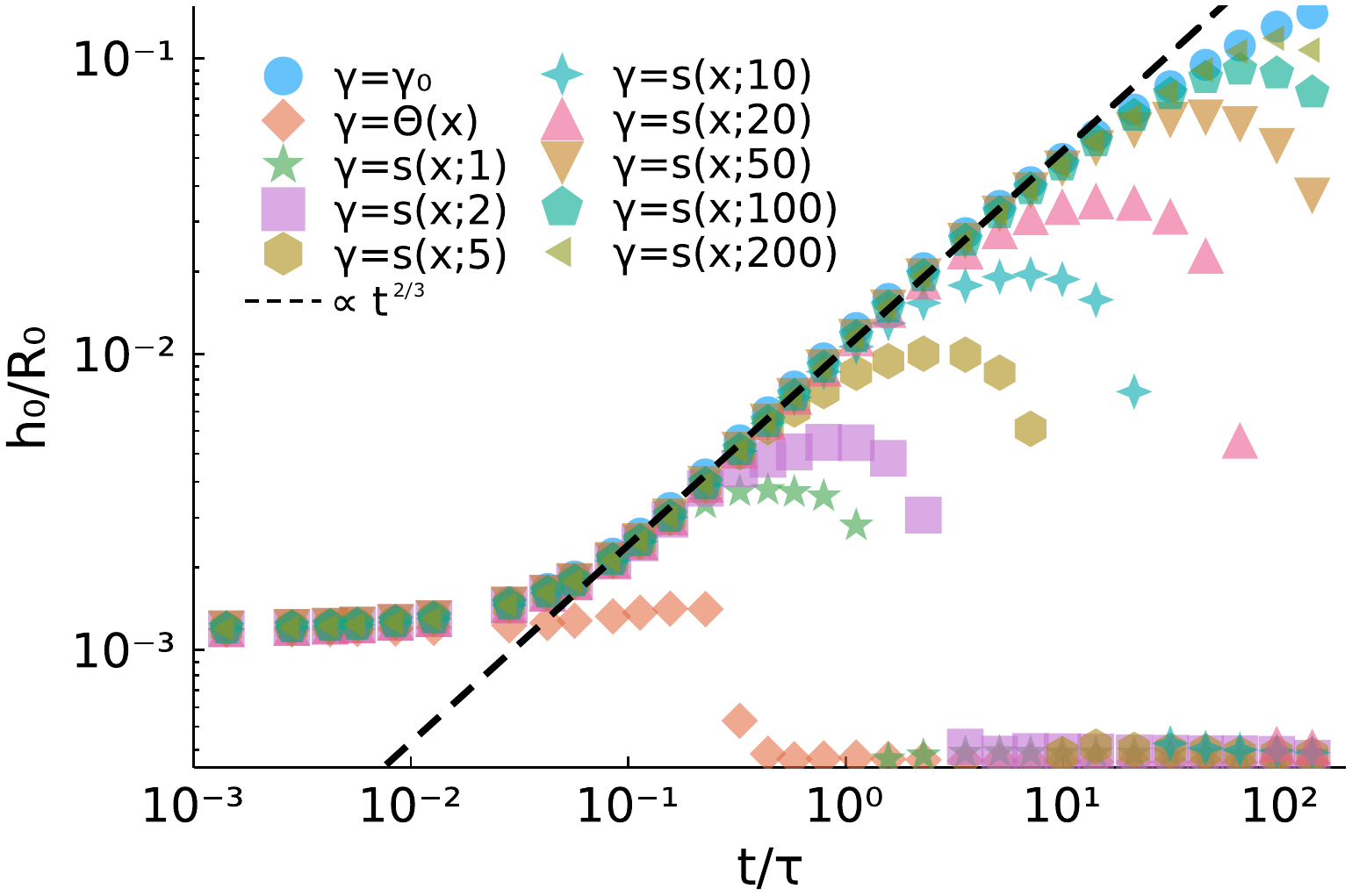}
    \caption{Time evolution of the liquid bridge $h_0(t)$ normalized with $R_0$. 
    Different symbols mark different surface tension gradients.
    The black dashed line displays the function $f(x) = \beta t^{2/3}$, with $\beta \approx 0.01$.}
    \label{fig:bridge_growth}
\end{figure}

The liquid bridge between two similar drops on a solid substrate with contact angle $\theta < \pi/2$ is expected to grow according to Eq.~(\ref{eq:coal_powerlaw}). 
Initially, this growth law has been derived for macroscopic drops for which the disjoining pressure is irrelevant.
The same growth rate holds true in our simulations, with $\Pi(h) \neq 0$ shown by the blue bullets in Fig~\ref{fig:bridge_growth}.
Only for late stages of the coalescence where $h_0 \sim h_d$ we observe a deviation from the black dashed line give by
\begin{equation}\label{eq:fit_powerlaw}
    f(t) = \beta t^{2/3},
\end{equation}
with $\beta = 0.01$.

If we use Eq.~(\ref{eq:gamma_step}) instead of Eq.~(\ref{eq:gamma_const}) the simulation yields the orange diamonds in Fig.~\ref{fig:bridge_growth}.
Clearly, the bridge does not grow, and the droplets do not coalesce.
Although we do not have a concentration field in our simulations, the results qualitatively agree with references~\cite{karpitschka2014sharp, doi:10.1021/la500459v, PhysRevLett.109.066103, doi:10.1021/la800630w}. 
Including $\Pi(h) \neq 0$, $h_0(t)$ even decreases, leaving only a thin precursor layer ($h\sim h_{\ast}$) between the droplets.
Instead of a steady motion of the twin droplet state, the droplet on the lower surface tension side ($\gamma = \gamma_0 - \Delta\gamma$) moves away from the step.
There are two effects to consider here, the Marangoni flow and the disjoining pressure.
The former keeps the bridge height small and the later tries to find an equilibrium for the droplets with $\theta$ as contact angle.
The moment $h\sim h_{\ast}$, the disjoining pressure dominates, and the result is a rearrangement of the three-phase contact line to have a stable droplet away from the surface tension gradient, shown by the orange curve in Fig~\ref{fig:final_state}.

Smearing out the surface tension gradient using Eq.~(\ref{eq:gamma_tanh}) is shown by the remaining symbols in Fig~\ref{fig:bridge_growth}.
Instead of an immediate separation of the droplets and independently of the smearing width $w$, the bridge starts to grow.
More surprisingly, the growth rate is in agreement with the power-law $t^{2/3}$ of Eq.~(\ref{eq:coal_powerlaw}).
While the bridge height grows, fluid flows into the region of higher surface tension, see Fig.~\ref{fig:drop_diff}.
The larger the bridge height, the easier it is for the Marangoni flow to drain liquid from the low surface tension droplet.
Eventually, the bridge height reaches a maximum and decreases rapidly to $h_0(t) \approx h_{\ast}$, similar to the step function.

\subsection{Separation and asymmetric coalescence}\label{subsec:separation}
\begin{figure}
    \centering
    \includegraphics[width=0.48\textwidth]{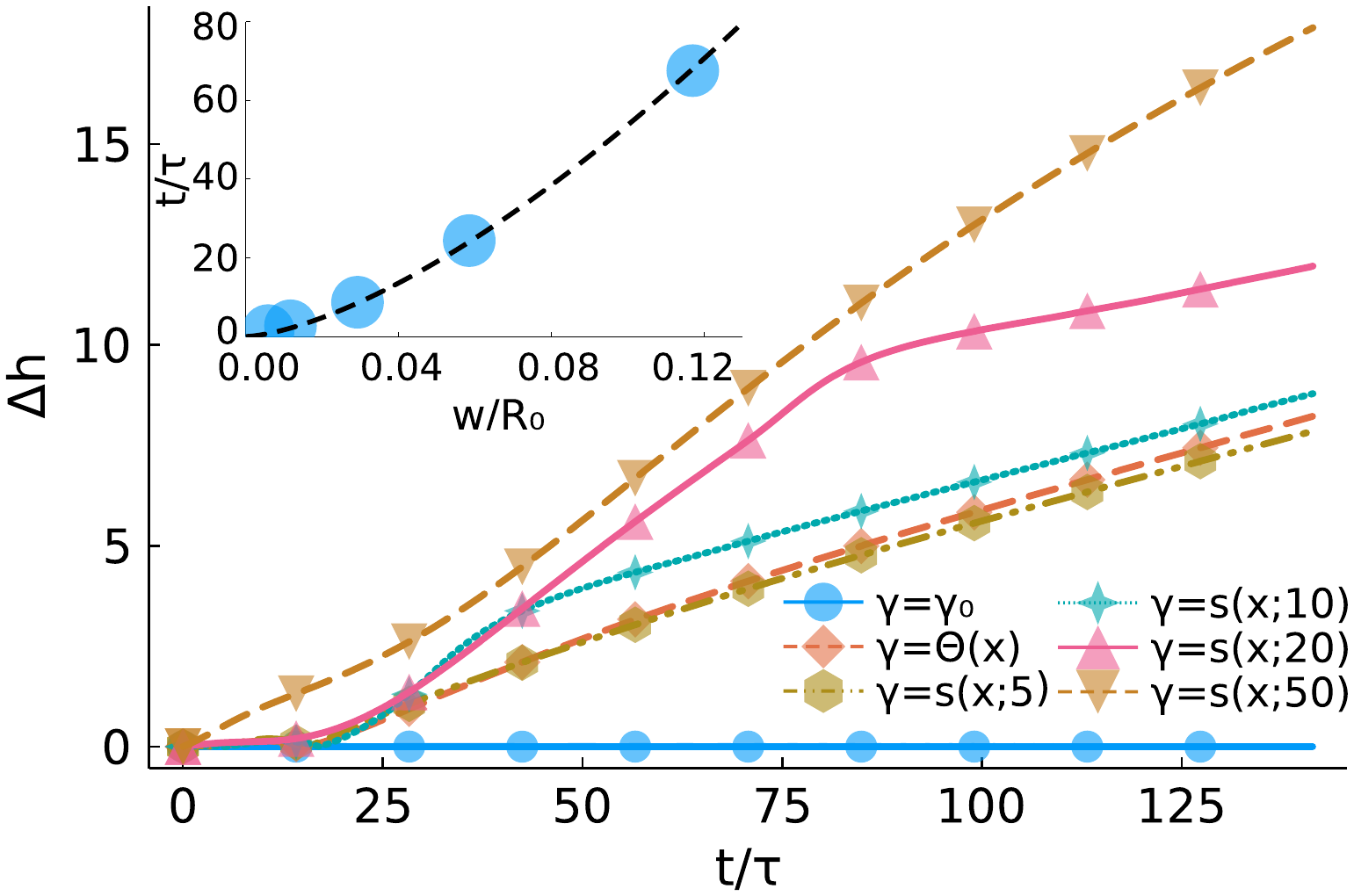}
    \caption{Difference between the maxima of the two droplets.
    In blue, we supply the symmetric coalescence using $\gamma(x) = \gamma_0$.
    The other curves are subject to a surface tension gradient, with different symbols and colors show different $\gamma(x)$.
    The inset displays the droplet separation time (blue bullets) for a subset of surface tensions.
    The black dashed line is a power-law $\sim w^{3/2}$.
    }
    \label{fig:drop_diff}
\end{figure}
The evolution of the bridge and therefore the flow is highly sensitive to the surface tension $\gamma(x)$, see Figs~\ref{fig:final_state},\ref{fig:bridge_growth}.
Choosing a constant surface tension $\gamma(x) = \gamma_0$, almost all data is in agreement with $\propto t^{2/3}$.
Given Eqs.~(\ref{eq:gamma_step}-\ref{eq:gamma_tanh}) the less smeared the transition in the surface tension is, the sooner the bridges deviate from the power-law.
Turning the argument around, the more space the transition between $\gamma_0$ and $\gamma_0-\Delta\gamma$ can occupy, the longer the growth rate agrees with $\propto t^{2/3}$.
Two things are happening, the first is that the bridge is moving.
The point of minimal thickness between the droplets shifts its location towards smaller surface tension.
While fluid is flowing in the direction of higher surface tension, the bridge is therefore travelling downstream.

This behavior is more pronounced for larger smearing values $w$.
Increasing $w$ yields smaller $\partial_x\gamma$ values at the bridge, as can be computed using Eq.~(\ref{eq:gamma_tanh}). 
For $w\gtrsim R_0$ we no longer see a separation, but an asymmetric coalescence.
A simpler measure to assess this statement is the difference in droplet height,
\begin{equation}\label{eq:drop_diff_h}
    \Delta h = \abs{h_{d,1} - h_{d,2}}
\end{equation}
where $h_{d,i}$ with $i=1,2$ identifies the left and right droplet, respectively.
These measurements are shown in Fig.~\ref{fig:drop_diff} for a subset of chosen surface tension fields.
For $w < 50$ (or $w/R_0 \le 1/3$) we observe a collapse of the data on short time scales.
The same curves, leaving aside the constant surface tension, show a kink in their slope at what we call the separation time $\tau_s$, which is defined according to $\min_t(h_0(\tau_{s}) \le h_{\ast})$ and shown in the inset of Fig~\ref{fig:drop_diff}.
On top of our data, the blue bullets, we plot a power-law $t \sim w^{3/2}$.
The motivation for this relation is as follows
\begin{equation}\label{eq:velsim1}
    \frac{\gamma}{\mu} \sim \frac{h_0}{2\mu}\partial_x\gamma,
\end{equation}
where $\gamma/\mu$ is a fluid specific capillary velocity and the left hand side is due to Marangoni~\cite{PhysRevLett.95.164503, doi:10.1021/la971292t}.
Rearranging the above similarity to have the bridge height on the left yields
\begin{equation}\label{velsim2}
    h_0 \sim \frac{2w\gamma}{\Delta\gamma},
\end{equation}
where we used $\partial_x\gamma \sim \Delta\gamma/w$, see Eq.~(\ref{eq:gamma_tanh}).
From Fig.~\ref{fig:bridge_growth} we know that Eq.~(\ref{eq:coal_powerlaw}) holds true for most of our data, therefore
\begin{equation}\label{velsim3}
    \tau_s \sim \sqrt{\frac{8\gamma^3}{(\Delta\gamma)^3}w^3}.
\end{equation}
The black dashed line in the inset of Fig.~\ref{fig:drop_diff} is given by $\sim w^{3/2}$, showing good agreement with the data.

\begin{figure}
    \centering
    \includegraphics[width=0.48\textwidth]{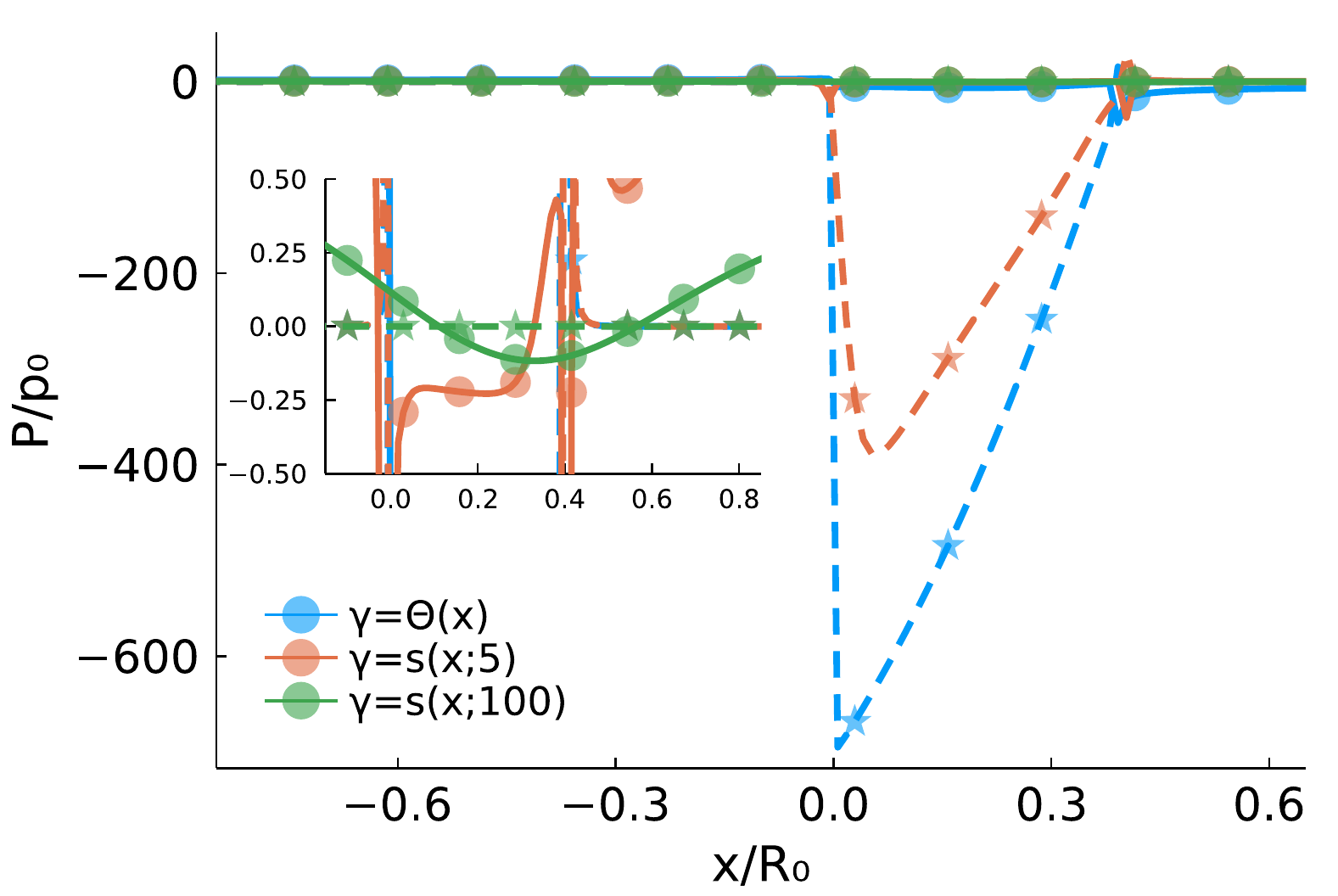}
    \caption{Pressure distribution around the bridge for three different $\gamma(x)$, Eq.~(\ref{eq:pressure}) at $t=140\tau$.
    Solid lines with bullets display $\partial_x^2 h$ while dashed lines with stars show $\Pi(h)$, Eq.~(\ref{eq:disjoin}.
    On the y-axis we normalize the pressures with $p_0$, see Eq.~\ref{eq:pressure_norm}.
    In the inset, we zoom into the region at $P/p_0 \approx 0$. 
    For non-separating cases, e.g. $\gamma(x) = s(x;100)$ (green curves) the disjoining pressure is vanishing.}
    \label{fig:pressures}
\end{figure}
To have a better understanding of this process, we measure the two pressure components $\partial_x^2 h$ and $\Pi(h)$, as shown in Fig.~\ref{fig:pressures}.
The full (dashed) lines with bullets (stars) show $\gamma(x)\partial_x^2 h$ ($\Pi(h)$). 
Different colors refer to different surface tensions, with all data being from the same time step at $t\approx 130\tau$.
To normalize both pressure components, we use 
\begin{equation}\label{eq:pressure_norm}
    p_0 = \frac{\gamma}{R},
\end{equation}
with $\gamma = \gamma_0$ and $R = R_0$.
Knowing that the coalescence is driven by capillarity ($\gamma\partial_x^2 h$) we expect that $\partial_x^2 h > \Pi(h)$ for $\gamma(x) = \gamma_0$.
This holds true not only for the constant surface tension for but also for large smearing $w\ge 100$, which is shown in the inset of Fig.~\ref{fig:pressures}.
Reducing the smearing and having a sharp surface tension transition, the disjoining pressure becomes the dominating pressure contribution.
While the value is large, the thickness of the precursor layer only allows for a limited flow. 
Still, we argue that at small scales ($R_0 \approx 1\mu m$) the disjoining pressure does influence the coalescence behavior quite significantly.
Lastly, the coalescence or non-coalescence is depending on the specific surface tension gradient and not only on the absolute surface tension difference ($\gamma_0-\Delta\gamma$).
We identify two scenarios for one $\Delta\gamma$, which are either droplet separation, yellow hexagons Fig.~\ref{fig:bridge_growth} and green hexagons Fig.~\ref{fig:drop_diff} or asymmetric coalescence, see cyan pentagons Fig.~\ref{fig:bridge_growth}.

\section{Summary and Conclusions}\label{sec:sum_conclu}

To summarize, we have performed numerical experiments of sessile droplet coalescence using a lattice Boltzmann method based on the thin film equation with an effective Marangoni contribution.
The droplets were subject to a spatially resolved surface tension gradient and a disjoining pressure $\Pi(h)$.
Making strong assumptions on the theoretical formulation, i.e. quasi static profile, we have shown that the disjoining pressure contribution affects the flow.
If there is, however, no surface tension gradient, the disjoining pressure does not alter the known scaling law for the bridge growth $h_0(t) \sim t^{2/3}$.

Using a spatially varying surface tension, we observe two different scenarios.
From reference experiments at larger scales with different but miscible liquids, we know that the Marangoni flow can prevent coalescence.
This behavior translates to the first scenario, where the droplets separate.
Using a step function between the two plateau values of $\gamma$, the bridge height quickly decreases, and the coalescence process is stopped.
Smearing the surface tension step using a tangent hyperbolicus we have a transient bridge growth that for small smearing widths $w$ leads to droplet separation.
The time it takes to separate the two droplets depends strongly on the surface tension gradient.
This dependence can be motivated using a balance of velocities, which results in the scaling $\tau_s\sim w^{3/2}$.  

If the smearing width becomes large ($\approx R_0$) the resulting flow promotes an asymmetric coalescence.
Similar to the constant surface tension, the product is a single droplet.
In contrast to a constant surface tension, the center of mass of this droplet is clearly shifted towards a region of higher surface tension.

In future work, we plan to perform simulations at larger scales using the volume of fluid method.
Therefore, having numerical experiments at scales where the disjoining pressure can safely be neglected.
Using a similarly well resolved surface tension gradient, we assume to confirm the case of asymmetric coalescence at larger scales, which in turn could be used for mixing purposes and additive printing processes.

\begin{acknowledgements}
S. Z., J. R. and K. M. acknowledge financial support from the Independent Research Fund Denmark through a DFF Sapere Aude Research Leader grant (grant number 9063-00018B).

\end{acknowledgements}

\appendix

\section{Thin film lattice Boltzmann method}\label{app:two}
This approach is based on the evolution of discrete probability density functions ($f_i$) where
\begin{equation}\label{eq:LBE}
    \begin{split}
        &f_i(x+c^{(i)}\Delta t,t+\Delta t) = \\
        &\left(1 - \omega\right) f_i(x,t) + \omega f_i^{(eq)}(x,t) + w_i \frac{\Delta t}{c_s^2} c^{(i)} F,
    \end{split}
\end{equation}
where $\omega = \Delta t/\tau_{\ast}$ is a relaxation frequency and $\tau_{\ast}$ is relaxation time. 
The total force acting on the fluid is collected in $F$.
We adopt the standard D1Q3 (one-dimensional) scheme with $3$ lattice velocities~\cite{krueger2017}, given by
\begin{equation}\label{eq:speeds}
c^{(i)}  = [0, c, -c], \quad i = 0, 1, 2,
\end{equation}
where the lattice speeds $c=\frac{\Delta x}{\Delta t}$, with weights
\begin{equation}
w_0 = \frac{2}{3},\quad w_{1,2} = \frac{1}{6},
\end{equation}
and the upper bound for information transport, the speed of sound $c_s^2=\frac{c^2}{3}$.
The equilibrium distribution functions $f_i^{(eq)}$ read~\cite{VANTHANG20107373}:
\begin{gather}
    f_{0}^{eq} = h\left(1-\frac{1}{2c^2}gh - \frac{1}{c^2}u^2\right),\nonumber\\
    f_{1}^{eq} = h\left(\frac{1}{4c^2}gh + \frac{1}{2c}u + \frac{1}{2c^2}u^2\right)\label{eq:equilibria},\\
    f_{2}^{eq} = h\left(\frac{1}{4c^2}gh - \frac{1}{2c}u + \frac{1}{2c^2}u^2\right),\nonumber
\end{gather}
where $g$ is the gravitational acceleration that enforces the hydrostatic pressure condition. 
Given the size of the droplet we are interested in, we neglect gravity and set $g=0$ for the remainder of this work.

The film thickness $h$ and the velocity at the free surface $u$ are moments of the distribution functions $f_i$~\cite{Salmon:1999:0022-2402:503, PhysRevE.65.036309, PhysRevE.104.034801}:
\begin{equation}\label{eq:hydrofields}
    h= \sum_{i=0}^2 f_i \qquad hu = \sum_{i=0}^2 c^{(i)} f_i.
\end{equation}
The force $F$ in (\ref{eq:LBE}) accounts for three terms,
\begin{equation}\label{eq:force}
    F = F_{\text{cap}} + F_{\text{fric}} + F_{\gamma}.  
\end{equation}
The first is the effect of the film pressure $p$, Eq.~(\ref{eq:pressure}), 
\begin{equation}\label{eq:capillary_force}
    F_{\text{cap}} = -\frac{1}{\rho_0} h \frac{\partial p}{\partial x},
\end{equation}
where $\rho_0$ is the fluid density. 
Viscous friction with the substrate is contained in
\begin{equation}\label{eq:fric_force}
    F_{\text{fric}} = -\nu \alpha_{\delta}(h) u,
\end{equation}
where $\nu=\mu/\rho_0$ is the fluid kinematic viscosity (related to the relaxation time $\tau$ by $\nu = c_s^2\left(\tau-\frac{\Delta t}{2}\right)$).
The thickness dependent function $\alpha_{\delta}(h)$ is approximately an inverse mobility $M(h)$,
\begin{equation}\label{eq:fric_alpha}
     \alpha_{\delta}(h) = \frac{6 h}{2h^2 + 6h\delta + 3\delta^2},
\end{equation}
with a slip length $\delta$.

\section{Lattice Boltzmann parameters}\label{app:one}
The grid size is $L=1024\Delta x$ with periodic boundary conditions except for $\partial_x\gamma(0) = \partial_x\gamma(L) = 0$.
Both droplets have a base radius $R_0 = 171\Delta x$, a maximum thickness $h_d = 30.15$ and form a contact angle $\theta = \pi/9$ with the substrate.
The initial bridge height is measured to be $h_0(0) \approx 0.2$.
The LBM relaxation time is set to unity, which results in a viscosity $\mu = 1/6$. 
We use a slip boundary condition, see Eq.~(\ref{eq:fric_alpha}), with $\delta \approx h_0/2$.
The surface tension $\gamma_0 = 10^{-5}$, while $\Delta\gamma = 0.2\cdot \gamma_0$.
In the disjoining pressure, Eq.~(\ref{eq:disjoin}) we set $h_{\ast} = 0.09$ and $(n,m) = (9,3)$.
The intertio-capillary time scale $\tau$ is given by $\tau \approx 7\cdot 10^5 \Delta t$.

\end{document}